\documentclass[twocolumn,amsmath,amssymb,aps,prx]{revtex4-2}

\usepackage{url}

\usepackage{blindtext}
\usepackage{bm}
\usepackage{amsmath}
\usepackage{color}
\usepackage{array}
\usepackage{tabularx}
\usepackage{multirow}
\usepackage[export]{adjustbox}

\usepackage{subfigure}
\usepackage{ulem}
\usepackage{xcolor}
\usepackage{graphicx}% Include figure files
\usepackage{dcolumn}% Align table columns on decimal point
\usepackage{bm}% bold math
\usepackage{hyperref}% add hypertext capabilities
\usepackage[mathlines]{lineno}% Enable numbering of text and display math
\usepackage{academicons}
\def\_#1{{\bf #1}}
\def\D{\nabla}
\def\.{\cdot}
\def\x{\times}
\def\Es{\_E_{\rm s}}
\def\Hs{\_H_{\rm s}}
\def\Ed{\_E_{\rm d}}
\def\Hd{\_H_{\rm d}}
\def\Et{\_E_{\rm t}}
\def\Ht{\_H_{\rm t}}
\def\rv{\_r}
\def\rvs{\_r_{\rm s}}
\def\rvd{\_r_{\rm d}}
\def\xs{x_{\rm s}}
\def\ys{y_{\rm s}}
\def\rs{r_{\rm s}}
\def\xd{x_{\rm d}}
\def\yd{y_{\rm d}}
\def\rd{r_{\rm d}}

\begin{document}

\preprint{APS/123-QED}

\title{Subwavelength Focusing by Engineered Power Flow-Conformal Metamirrors}% Force line breaks with \\

\author{Hamidreza Taghvaee\href{https://orcid.org/0000-0001-8732-6086}{\includegraphics[scale=0.5]{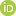}}$^{1,3,4}$}
 \email{hamidreza.taghvaee@upc.edu}
\author{Fu Liu\href{https://orcid.org/0000-0002-7393-2963}{\includegraphics[scale=0.5]{./ORCID-iD_icon-16x16.png}}$^{2,3}$}
\author{Ana Díaz-Rubio\href{https://orcid.org/0000-0002-0115-1834}{\includegraphics[scale=0.5]{./ORCID-iD_icon-16x16.png}}$^{3}$}
%\author{Sergi Abadal\href{https://orcid.org/0000-0003-0941-0260}{\includegraphics[scale=0.5]{./figures/ORCID-iD_icon-16x16.png}}$^{1}$}
%\author{Albert Cabellos-Aparicio$^{1}$}
%\author{Eduard Alarc\'{o}n$^{1}$}
\author{Sergei Tretyakov\href{https://orcid.org/0000-0002-4738-9987}{\includegraphics[scale=0.5]{./ORCID-iD_icon-16x16.png}}$^{3}$}

\affiliation{$^{1}$NaNoNetworking Center in Catalonia (N3Cat), Universitat Polit\`{e}cnica de Catalunya, 08034 Barcelona, Spain.}
\affiliation{$^{2}$Key Laboratory for Physical Electronics and Devices of the Ministry of Education and Shaanxi Key Lab of 
Information Photonic Technique, School of Electronic Science and Engineering, Faculty of Electronic and Information Engineering, Xi'an Jiaotong University, Xi'an 710049, China}%
\affiliation{$^{3}$Department of Electronics and Nanoengineering, Aalto University, P.O. Box 15500, FI-00076 Aalto, Finland}
\affiliation{$^{4}$School of Mathematical Sciences, Department of Electrical and Electronics Engineering, University of Nottingham, Nottingham NG7 2RD U.K.}

%\author{Ana Díaz Rubio}
%\affiliation{
%\affiliation{%


\date{\today}% It is always \today, today,

\begin{abstract}
Many advances in reflective metasurfaces have been made during the last few years, implementing efficient manipulations of wavefronts, especially for plane waves.  Despite numerous solutions that have been developed throughout the years, a practical method to obtain subwavelength focusing without the generation of additional undesired scattering is a challenge to this day. In this paper, we introduce and discuss lossless reflectors for focusing incident waves into a point.  The solution is based on the so-called power flow-conformal surfaces that allow theoretically arbitrary shaping of reflected waves. The metamirror shape is adapted to the power flow of the sum of the incident and reflected waves, allowing a local description of the reflector surface based on the surface impedance. In particular, we present a study of two scenarios. First, we study the scenario when the field is emitted by a point source and focused at an image point (in the considered example, with the $\lambda/20$ resolution). Second, we analyze a metasurface capable to focus the power of an illuminating plane wave. This work provides a feasible strategy for various applications, including detecting biological signals near the skin, sensitive power focusing for cancer therapy, and point-to-point power transfer.

%\begin{description}
%\end{description}
\end{abstract}

\maketitle

%\tableofcontents

\section{Introduction}
\label{sec:intro}
The development of lenses and focusing reflectors as optical tools started in ancient times. Enabling and improving the vision of small objects intrigued the human mind and led to the invention of many optical instruments. However, the wave nature of light limits the quality of focusing, as was explained by Ernst Abbe in 1873 \cite{Abbe1873,Abbe}. In astronomy, the diffraction-limited angular resolution of a telescopic lens is inversely proportional to the diameter of the aperture, $D$, and proportional to the wavelength of the light being observed, $\lambda$, as $1.22\lambda/D$ \cite{Rayleigh, Wang2015}. In microscopy, this limit impedes the resolution of imaging roughly around half-wavelength \cite{Gray2009}. It also limits the ability to focus waves into a small spot (hotspot), which is important, in particular, for photolithography \cite{Paik2020}.

In the modern era, methods for focusing and concentration of electromagnetic energy in small subwavelength regions are crucial in a variety of applications including therapy \cite{23024}, energy harvesting \cite{8892014}, wireless power transfer \cite{6175400}, and particle manipulation \cite{Yuksel2020}. 
The problem arises because scattered waves from an object that carry components with high spatial frequencies correspond to evanescent waves that exponentially decay in space. There are several known approaches for realizing subwavelength focusing in the far zone. Let us briefly mention the main methods. 

One is based on the use of artificial materials (metamaterials) with simultaneously negative permittivity and permeability, discussed by V. Veselago in 1968  \cite{veselago}.  Later, in 2000, J. Pendry showed that a double-negative material slab theoretically functions as a perfect lens  \cite{PhysRevLett.85.3966}. However, limitations due to material losses and discrete structure of metamaterials do not allow practical realizations, see, e.g., \cite{5062500,Kim2015}. Using silver as a natural near-field optical superlens for evanescent modes only, the diffraction limit can be reduced down to approximately $\lambda/6$, however, due to dissipative losses, power is highly attenuated \cite{1536712,Fang534,Zhang2008,4964498}. 

Another approach is the use of extremely anisotropic materials, that allow conversion of evanescent modes into propagating ones and this way transporting images with subwavelength resolution to electrically long distances. For this purpose, wire media \cite{PhysRevB.71.193105,PhysRevB.73.073102,PhysRevB.73.033108} and hyperbolic materials \cite{Liu1686,Kildishev:07,doi:10.1063/1.3211115,Rho} have been used.

%\textcolor{green}{Metasurfaces, classical design} 
Alternative solutions have been proposed using metasurfaces, a two-dimensional version of metamaterials. 
%The main idea is to engineer the local reflection/transmission produced by each inclusion of the metasurfaces in such a way that the scattered fields add in phase at the  focal point  \cite{Yang2017,Liang}.
%Thin layer metamaterials or metasurfaces predicted to substantially enhance evanescent waves, compensating for the evanescent loss outside the component with high efficiency after all, their ability of focusing is not promising \cite{Liang}. 
One possibility is to use a double array of small resonant particles. Resonant oscillations in two parallel arrays that are strongly coupled realize the same effect of resonant amplification of evanescent waves as in the Veselago-Pendry lens based on bulk double-negative materials \cite{doi:10.1063/1.1765865,doi:10.1063/1.1922074,PhysRevB.74.235425,https://doi.org/10.1002/mop.22689}.
Using nonlinear metasurfaces, focusing and light concentration can be in principle realized reversing the phase of the propagating wave spectrum \cite{doi:10.1063/1.1604935,Maslovski_2012,Pendry71}.

Furthermore, there are works on the use of zone plates for evanescent modes \cite{Marks:05,Merlin927} and near-field plates \cite{4636830,Grbic511}. These structures, however, can be used only if the object is located in the vicinity of the plate (much smaller than the wavelength). This disadvantage can be overcome with the use of super-oscillations \cite{doi:10.1021/nl9002014,22447113,Yuan2017}, but at the cost of scattering most of the incident power into parasitic propagating modes. 

Despite the broad variety of methods known to allow subwavelength field concentration and focusing, all of them have their specific limitations and disadvantages, and it is important to explore other possibilities that may offer complementary advantages and further develop our understanding of these phenomena. 

%We note that in many cases the achieved subwavelength resolution relies on numerical optimizations, e.g., \cite{Yang2017},     but often lacks a well-developed theory. For instance, in \cite{David:15}, a specific layout of a planar metal–oxide–silicon is introduced but it is not justified from a theoretical point of view. Although paper \cite{Chang2018} proposes layouts with different nanoapertures, the methodology is not well-established.

In this work, we are interested in forming subwavelength focal spots in free space or in the presence of a small receiver (drain) using reflectors. 
%study reflective metasurfaces, also known as metamirrors, to implement highly efficient superlenses. 
We explore subwavelength focusing possibilities offered by recently introduced power flow-conformal metamirrors.  
%These metamirrors can perform as a highly efficient superlens, with  the ability to confront the fundamental problem of diffraction limit. 
Power flow-conformal reflectors have been recently proposed and applied to realize anomalous reflection and beam splitting   \cite{Rubioeaau7288,diaz2020dual} and for the design of refractive focusing devices \cite{peng2021efficient}
%Here, we extend the concept to a more advanced and demanding functionality: subwavelength focusing. 
The main operating principle of power flow-conformal metamirrors is to define the surface profile of the metamirror to be tangential to the power flow of the desired field distribution.
%Power flow analysis allows us to plant subwavelength metamirrors (as the building blocks of our design) along the direction of the Poynting vector.
When the reflector is shaped so that the power flow is at every point tangential to the surface, wave power flow does not cross the metasurface boundary and, consequently, we can locally define the properties of the metamirror as a lossless impedance boundary. In this case, the response of the metamirror can be modeled by local surface impedance $Z_{\rm s}$ (the ratio between the tangential components of the electric and magnetic fields at the surface) and this value is purely imaginary ($\Re[Z_{\rm s}]=0$) at every point of the reflector. In practice, this simplifies the design and implementation, because the response is local and each element of the metamirror can be designed by appropriately controlling the phase reflection between $0$ and $2\pi$ \cite{Rubioeaau7288,diaz2020dual}.

In this paper, we study two-dimensional power-flow conformal metamirrors for subwavelength focusing. In particular, we assume that the desired field structure is the sum of the incident field and a cylindrical wave that converges to a point focus. We study two different focusing scenarios. First, we study the scenario when the field emitted by a point source propagates in space and focuses at the image point with theoretically perfect resolution. Second, we analyze a metasurface capable to focus the power from an illuminating plane wave. In both cases, the proposed methodology is evaluated through full-wave simulations, using the finite element method in COMSOL Multiphysics. The results show an example of subwavelength focusing with Half Power Beam Width (HPBW) or hot-spot size as $\lambda/20$. Also, almost all of the incident power is concentrated at the focal point. 

%The paper is organized as follows: Section~\ref{sec:cavity} describes the analytical process to design power flow-conformal metamirror, in general and for focusing. Starting from the analysis of anomalous reflector we present the methodology to focus the field emitted by a point source. Then, in Section~\ref{sec:open}, we study open metamirrors to focus the field of an impinging plane wave. Finally, Section~\ref{sec:con} concludes the paper.

\section{From power flow-conformal anomalous reflectors to focusing mirrors} 
\label{sec:cavity}

\begin{figure}[ht]
    \centering
   % \vspace{-1cm}
    %\includegraphics[width=\linewidth]{1.jpg}
    \subfigure[]{\includegraphics[height=4.7cm,keepaspectratio]{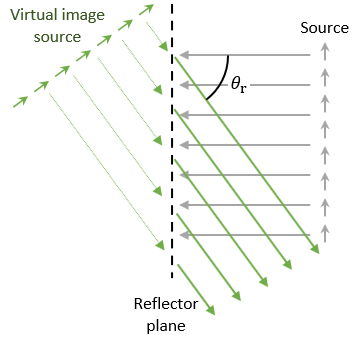} \label{fig:n1a}}
    \subfigure[]{\includegraphics[height=4.7cm,keepaspectratio]{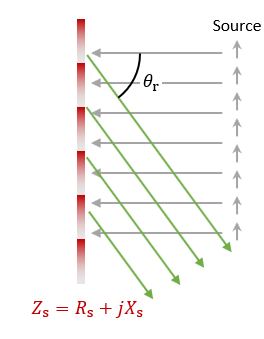} \label{fig:n1b}}
    \subfigure[]{\includegraphics[height=4.7cm,keepaspectratio]{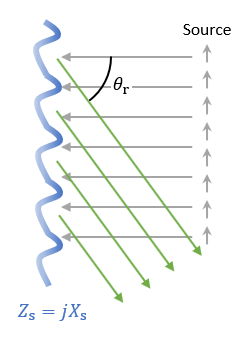} \label{fig:n1c}}
    \caption{Power flow-conformal anomalous reflectors. (a) Schematic representation of the problem. (b) A flat anomalous reflector is implemented by an ``active-lossy" metamirror. (c) Lossless power flow-conformal anomalous reflectors.}
    \label{fig:n1}
\end{figure}

\begin{figure*}[t]
    \centering
    \vspace{-1cm}
    \subfigure[]{\includegraphics[width=0.32\linewidth]{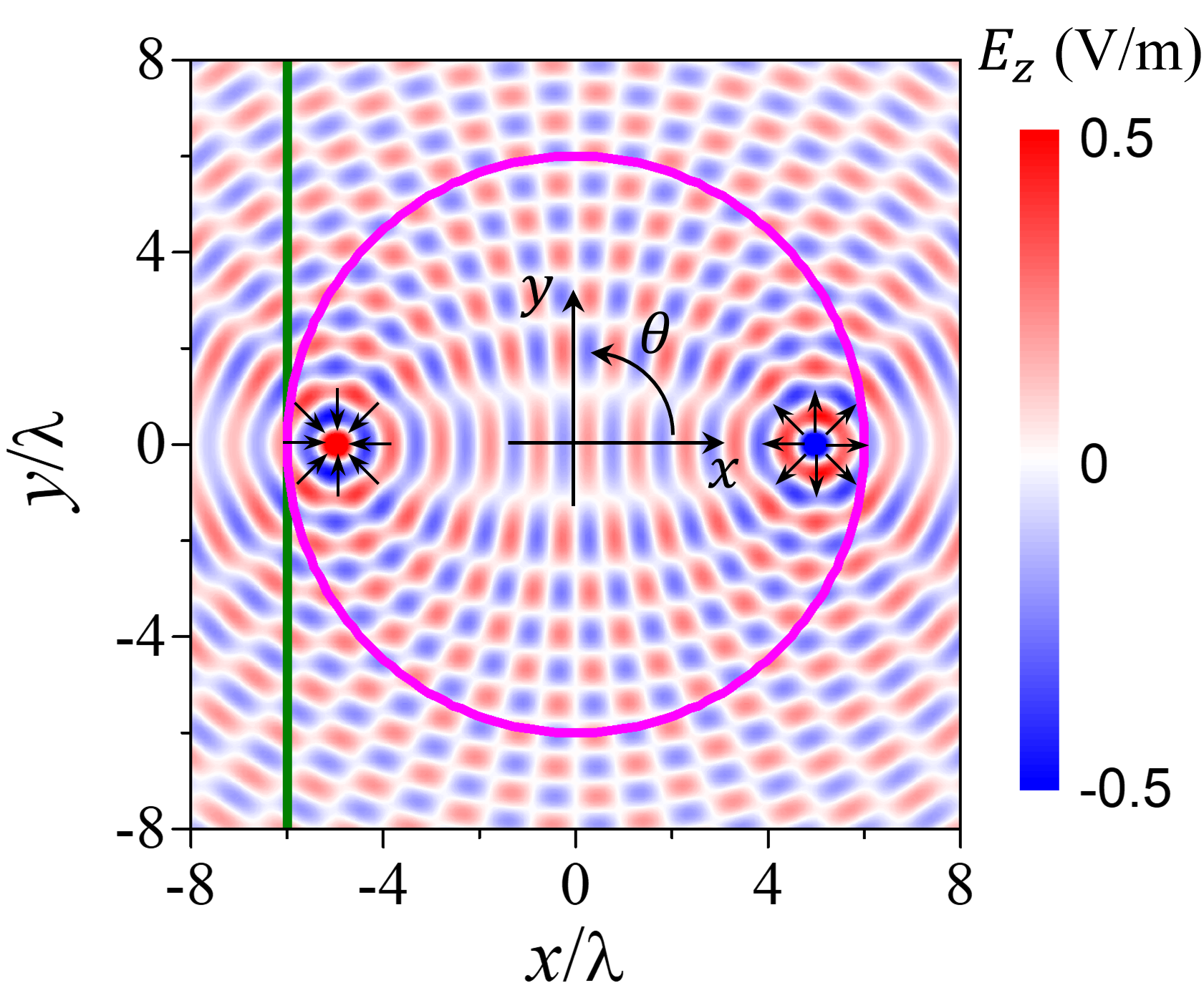}\label{fig:n2a}}
    \subfigure[]{\includegraphics[width=0.32\linewidth]{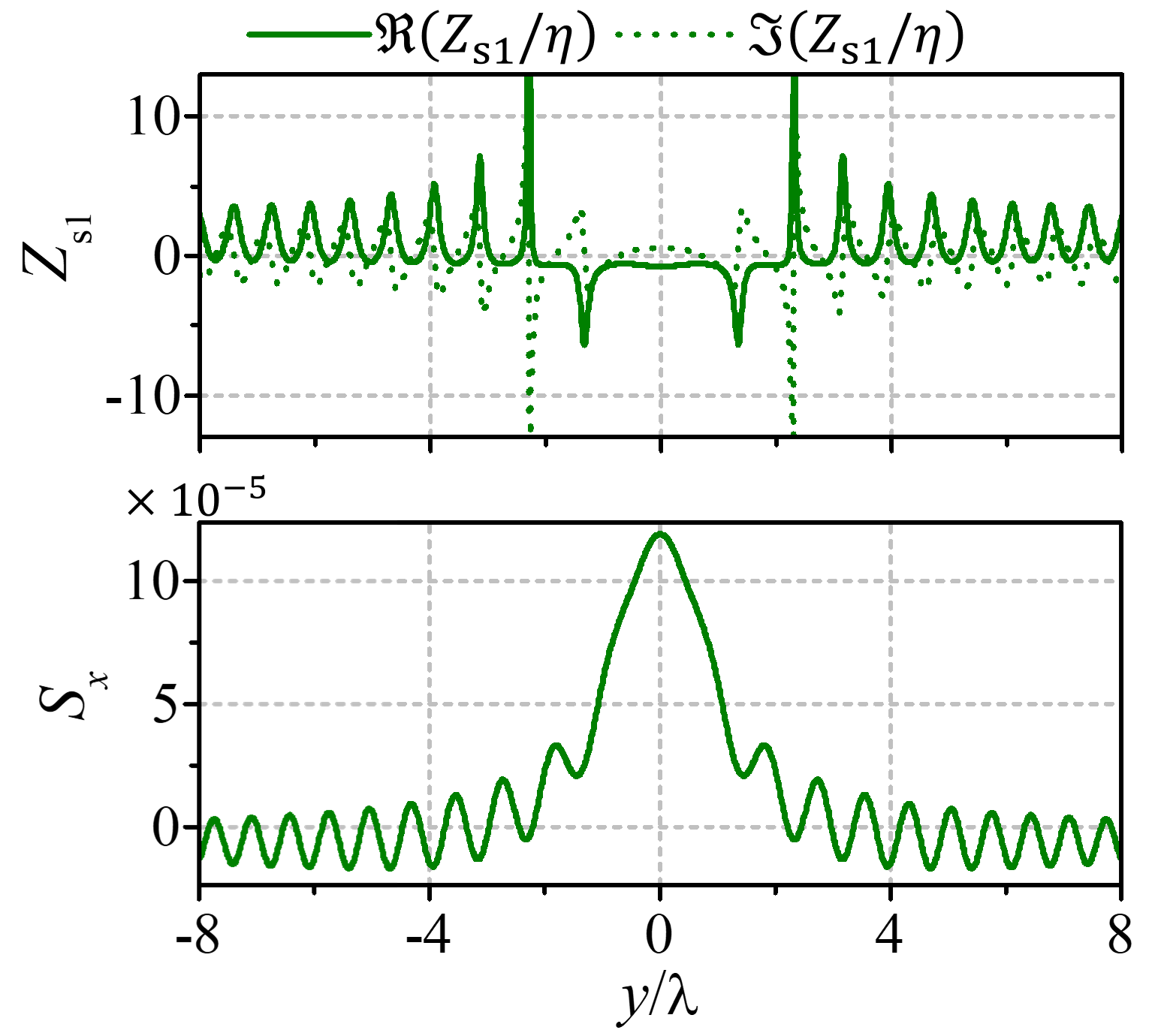}\label{fig:n2b}}
    \subfigure[]{\includegraphics[width=0.32\linewidth]{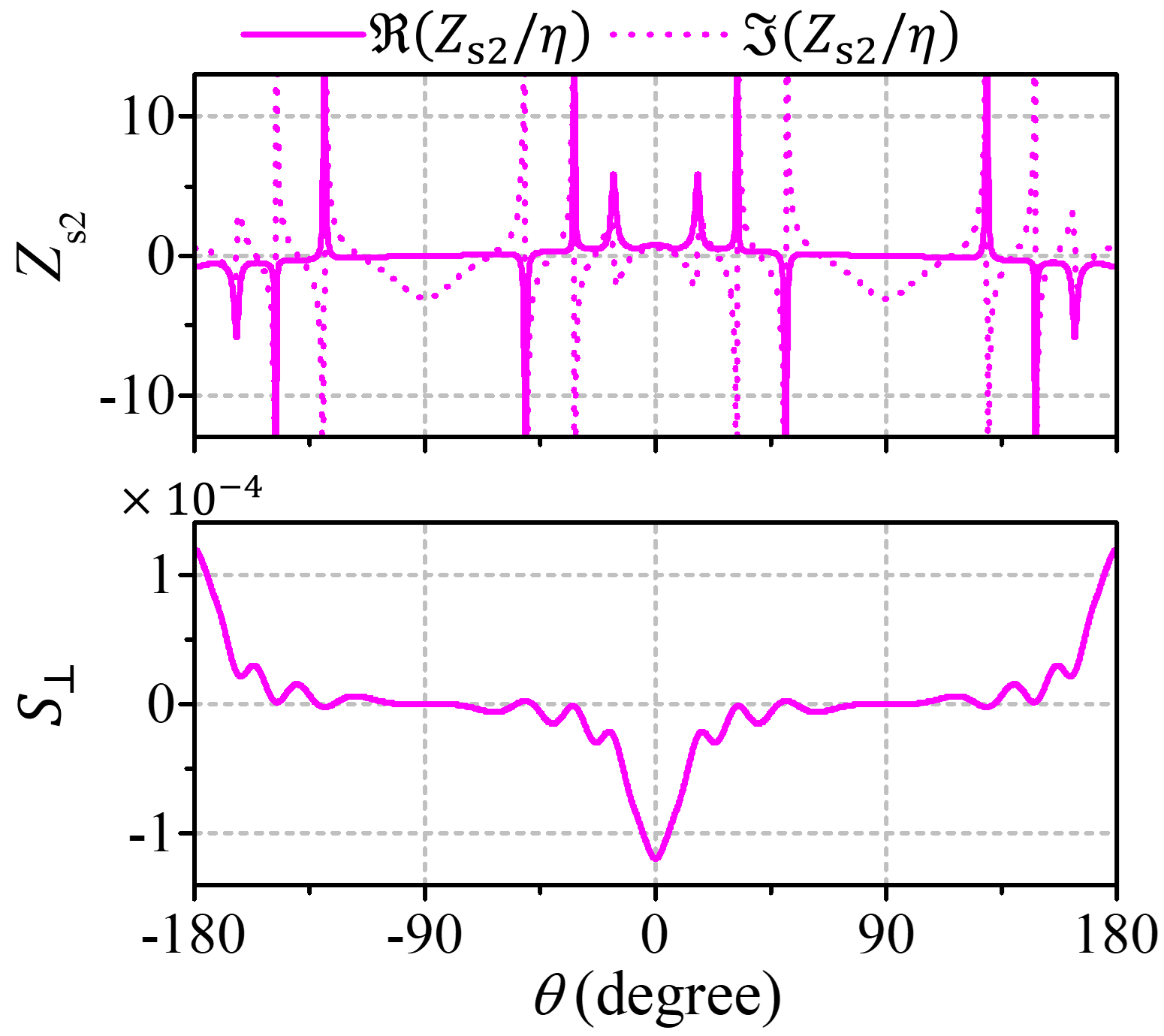}\label{fig:n2c}}
    %\subfigure[]{\includegraphics[width=0.32\linewidth]{2.jpg}}
    \caption{Fields of a point  source located at $\rvs=(5\lambda, 0)$  propagate  and  focus  at  the  image point,  located at $\rvd=(-5\lambda, 0)$. (a) The total electric field in the  $xy$-plane when the current source amplitude is $I_0=4/k\eta$. Black arrows represent the power emerging from the source and entering into the image point. The green and magenta lines represent two different surfaces for implementing the metamirror. (b) Electromagnetic properties of a flat metamirror along the $y$-direction located at $x=-6\lambda$. The upper panel shows the surface impedance at the flat surface. The bottom panel shows the local power entering and emerging from the metamirror. (c) Electromagnetic properties of a circular metamirror with the radius  $6\lambda$, centered at the coordinate origin.  The upper panel shows the surface impedance. The bottom panel shows the local power entering and emerging from the metamirror.}
    \label{fig:n2}
\end{figure*}

Before starting with the design of focusing mirrors, let us recall the general operating principle of power flow-conformal metamirrors by analyzing the anomalous reflector scenario.  Anomalous reflectors capable of reflecting the energy of an impinging plane wave into arbitrary direction have been intensively studied during the past years~\cite{Rubioeaau7288,diaz2020dual,PhysRevB.96.125409,diaz2017generalized,PhysRevLett.117.256103,ra2017metagratings}. Based on the uniqueness theorem, an impedance boundary condition is enough to define the desired power flow in front of a reflector. Thus, by engineering the surface impedance of a boundary we can enable desired functionalities. To understand the design and realization challenges, we can analyze an anomalous reflector that transforms a normally incident plane wave into a plane wave propagating in an oblique direction defined by the angle $\theta_{\rm r}$. We can form the desired field distribution assuming that the reflected plane wave is generated by a virtual source placed behind the reflector plane [see Fig.~\ref{fig:n1a}] whose amplitude is defined to warranty the power conservation (the component of the Poynting vector normal to the reference plane is the same for the incident and reflected waves). As it has been shown in~\cite{Rubioeaau7288,diaz2020dual,PhysRevB.96.125409,diaz2017generalized,PhysRevLett.117.256103}, if we define a flat surface as the reflector plane, the required surface impedance is defined as
\begin{equation}
\Et=	Z_s\hat{n}\times\Ht ,
	\label{Eq:Zs}
\end{equation}
where $\Et$ and $\Ht$ are the total electric and magnetic fields ($\Et$ only has a $z$-component), $\hat{n}$ is the normal unit vector pointing towards the incident wave source, and surface impedance $Z_s$ is a complex quantity whose real part takes positive and negative values. This is because the two plane waves of the desired field structure (the incident and reflected one) interfere.  The real part of the complex surface impedance represents power being absorbed when it is positive and power emerging from the surface when it is negative. This scenario is illustrated in Fig.~\ref{fig:n1b}
At this point, it is important to mention that, because the amplitude of the virtual source generating the reflected field has been adjusted to satisfy power conservation, the averaged power crossing the surface is zero meaning that the metamirror is overall lossless. 
One of the possibilities to realize perfect anomalous reflectors is to engineer the shape of the metamirror in such a way that the surface is always tangential to the spatial power flow generated by the two plane waves ~\cite{Rubioeaau7288,diaz2020dual}, see Fig.~\ref{fig:n1c}. In this case, the surface impedance of the metamirror is purely imaginary at all points, and the surface can be implemented using simple lossless meta-atoms. In what follows, we will apply this  methodology to the design of focusing devices. 

Now, let us consider the case of a cylindrical wave emitted from a line source (a point source in 2D space) that propagates in air (vacuum) and focuses on the image line with theoretically perfect resolution. We assume that the current line source is infinite along the $z$-direction, and therefore we only need to consider the wave propagation on the $x$-$y$ plane. For such a line source, the electric field of the emitted cylindrical wave only has a $z$-component, and the fields can be written as
\begin{equation}
	\Es = -\frac{\eta k I_0}{4}H_0^{(2)}(k\rs)\hat{z},\hspace{5mm}
	\Hs = \frac{j}{\eta k}\D\x\Es,
	\label{inc_w}
\end{equation}
%\begin{eqnarray}
%	\Es & = & \frac{\eta k I_0}{4}H_0^{(2)}(k\rs)\hat{z},\\
%	\Hs & = & \frac{j}{\eta k}\D\x\Es,
%\end{eqnarray}
%\begin{eqnarray}
%E_{{\rm s}z}=\eta\frac{ k}{4}I_0H_0^{(2)}(kr_{\rm s})\\
%H_{{\rm s}x}=j\frac{I_0}{4}\frac{\partial H_0^{(2)}(kr_{\rm s}) }{\partial_y}=-jI_0\frac{k}{4}\frac{y-y_{\rm s}}{r_{\rm s}}H_1^{(2)}(kr_{\rm s}) \\
%H_{{\rm s}y}=-\frac{j}{4}I_0\frac{\partial H_0^{(2)}(kr_{\rm s}) }{\partial_x}=jI_0\frac{k}{4}\frac{x-x_{\rm s}}{r_{\rm s}}H_1^{(2)}(kr_{\rm s}) 
%\label{eq:inc}
%\end{eqnarray}  
where the subscript $\rm{s}$ denotes source fields, $\eta$ is the free-space impedance, $k$ is the vacuum wavenumber at the operation frequency, $I_0$ is the amplitude of the source current, $H_0^{(2)}$ is the zero-th order Hankel function of the second kind with $\rs=|\rv-\rvs|=\sqrt{(x-\xs)^2+(y-\ys)^2}$ being the distance from the observation point $\rv=(x,y)$ to the current location $\rvs=(\xs,\ys)$, and $\hat{z}$ is the unit vector in $z$ direction. The harmonic time dependence in form $\exp(j\omega t)$ is assumed.

For perfect focusing, we desire the same amount of energy converging at a drain line. Thus, the desired field distribution in space is the sum of the incident cylindrical wave \eqref{inc_w} and a converging cylindrical wave 
\begin{equation}
	\Ed = \frac{\eta k I_0}{4}H_0^{(1)}(k\rd)\hat{z},\hspace{5mm}
	\Hd = \frac{j}{\eta k}\D\x\Ed,
	\label{E_sink}
\end{equation}
%\begin{eqnarray}
%E_{{\rm d}z}(x, y)=\eta\frac{ k}{4}I_0H_0^{(1)}(k r_{\rm d})\label{E_sink}\\
%H_{{\rm d}x}= \frac{j}{4}I_0\frac{\partial H_0^{(1)}(kr_{\rm d}) }{\partial_y}=-jI_0\frac{k}{4}\frac{y-y_{\rm d}}{r_{\rm d}}H_1^{(1)}(kr_{\rm d}) \label{H_dx}\\
%H_{{\rm d}y}=-\frac{j}{4}I_0\frac{\partial H_0^{(1)}(kr_{\rm d}) }{\partial_x}=jI_0\frac{k}{4}\frac{x-x_{\rm d}}{r_{\rm d}}H_1^{(1)}(kr_{\rm d})\label{H_dy}
%\label{eq:div}
%\end{eqnarray}  
where subscript $\rm{d}$ denotes the desired reflected fields and $H_0^{(1)}$ is the zero-th order Hankel function of the first kind with $\rd=|\rv-\rvd|$ being the distance from the observation point to the drain location $\rvd=(\xd,\yd)$.
The total fields read
\begin{equation}
	\Et = \Es+\Ed,\hspace{5mm}
	\Ht = \Hs+\Hd,
\end{equation}
and the Poynting vector of the total field is
\begin{equation}
	\_S=S_x\hat{x}+S_y\hat{y}=\frac{1}{2}\Re[\Et\x\Ht^*].\label{eq:Poyntingvector}
\end{equation}
As an example, Fig.~\ref{fig:n2a} represents the field generated by current line located at $(5\lambda,0)$ with the amplitude $I_0=4/k \eta$ that is focused at point $(-5\lambda,0)$. Black arrows represent the power flow emerging from the source and converging into the focal point. 

For the design of a metamirror capable of performing this transformation, one can consider a flat surface along $y$-direction. Following the same example presented in Fig.~\ref{fig:n2a}, we can define a flat surface located at $x=-6\lambda$, e.g., the vertical green line, and calculate the required surface impedance. As it is shown in Fig. ~\ref{fig:n2b}, the surface impedance is a complex value whose real part takes positive and negative values, meaning that the metamirror locally produces losses or gain. 
The local gain/loss introduced by the metamirror can be evaluated using the normal component of the Poynting vector to the surface.
Specifically, the vectorial components of the Poynting vector $S_x$ and $S_y$ can be expressed in terms of the total fields as in Eq.~(\ref{eq:Poyntingvector}).
%, $\vec{S}=S_x \hat{x}+S_y \hat{y}$,  can be expressed in terms of the total fields as 
%\begin{eqnarray}
%S_x=-\frac{1}{2}\Re[E_{{\rm tot}z} H_{{\rm tot}y}^*], \\
%S_y=\frac{1}{2}\Re[E_{{\rm tot}z} H_{{\rm tot}x}^*].
%\end{eqnarray}
In this particular example of a flat surface along $y$-direction located at $x=-6\lambda$, the local power density crossing the surface in $x$-direction is represented by $S_x$ in  Fig.~\ref{fig:n2b}. 
If we calculate the net power (average of the normal component of the Poynting vector along the surface) we can see that, for finite-size metasurfaces, the metamirror is active and contributes to the power collected into the drain. Only in the limit case when the metasurface is infinite losses and gain are compensated and the structure is overall lossless, and this is the case for any open surface.

An overall lossless metamirror can be constructed if it is placed on a closed surface enclosing both the source and the drain (which is a virtual source generating the reflected fields). As an example, we analyze a circular metamirror with the radius  $6\lambda$ and centered at the origin of the coordinate system [see the magenta circle in Fig.~\ref{fig:n2a}]. To find the required surface impedance, it is necessary to calculate the tangential components of the magnetic field at each point of the surface. To this end, we define the normal vector of the metamirror as $\hat{n}=n_x\hat{x}+n_y\hat{y}=-\cos\theta\hat{x}-\sin\theta\hat{y}$. The tangential component of the magnetic field reads $H_t=\hat{n}\times\Ht=H_{{\rm tot}y}n_x-H_{{\rm tot}x}n_y$. The surface impedance of the metamirror calculated with Eq.~\eqref{Eq:Zs} is represented in Fig.~\ref{fig:n2c}. In this case, the surface impedance is also a complex number with a non-zero real part, meaning that the metamirror is locally ``active/lossy". Figure~\ref{fig:n2c} shows the normal component of the Poynting vector $S_\perp=S_xn_x+S_yn_y$ at each point of the surface where we can see that the average value of the power crossing the surface is zero. Because any close boundary enclosing both the source and drain is overall lossless, we can apply the principles of power flow-conformal metamirrors to find a shape of a closed boundary that locally produces lossless behavior.

\begin{figure}[t]
    \centering
    %\vspace{-1cm}
    \subfigure[]{\includegraphics[width=0.7\linewidth]{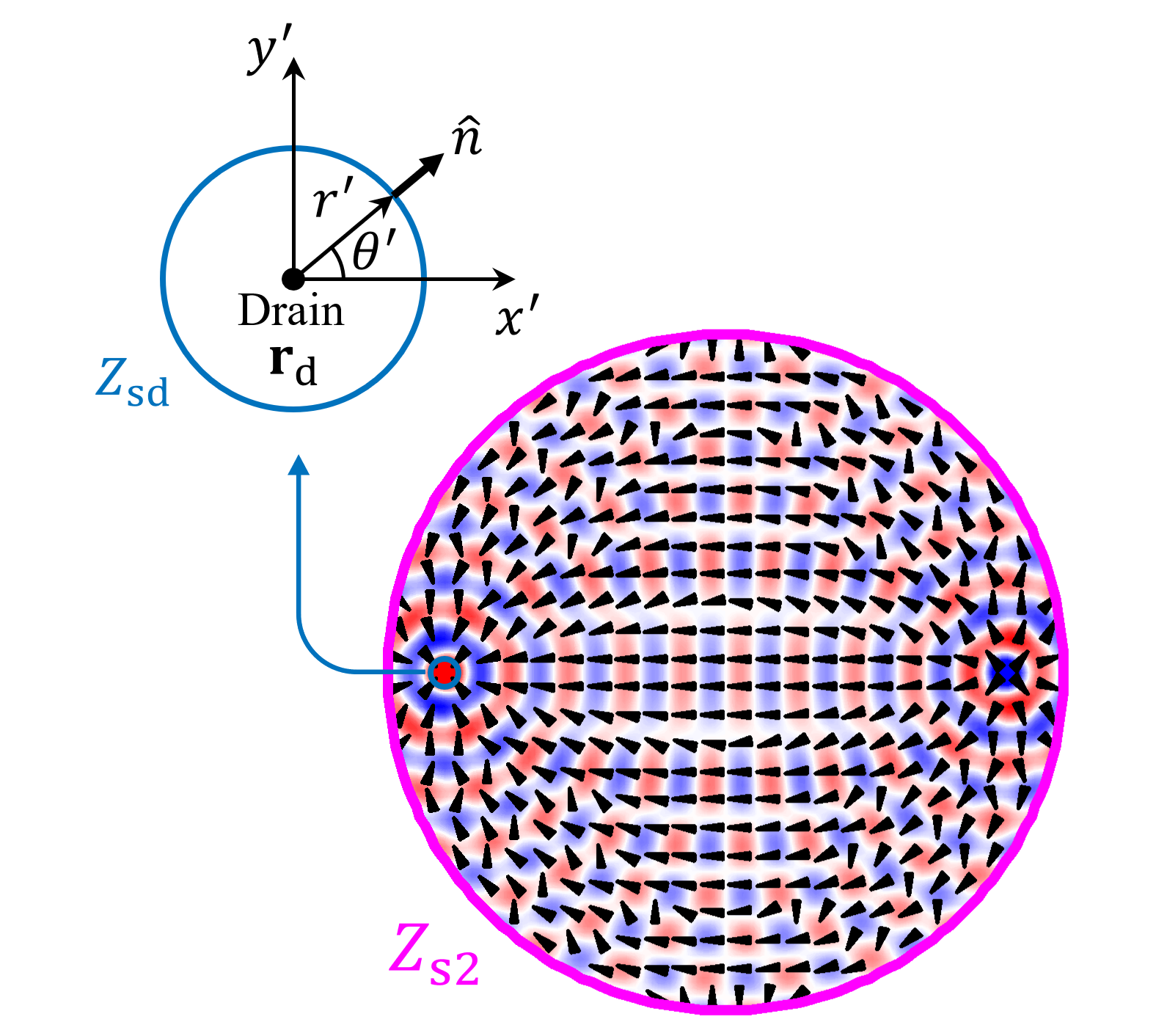}\label{fig:n3a}}
    \subfigure[]{\includegraphics[width=0.8\linewidth]{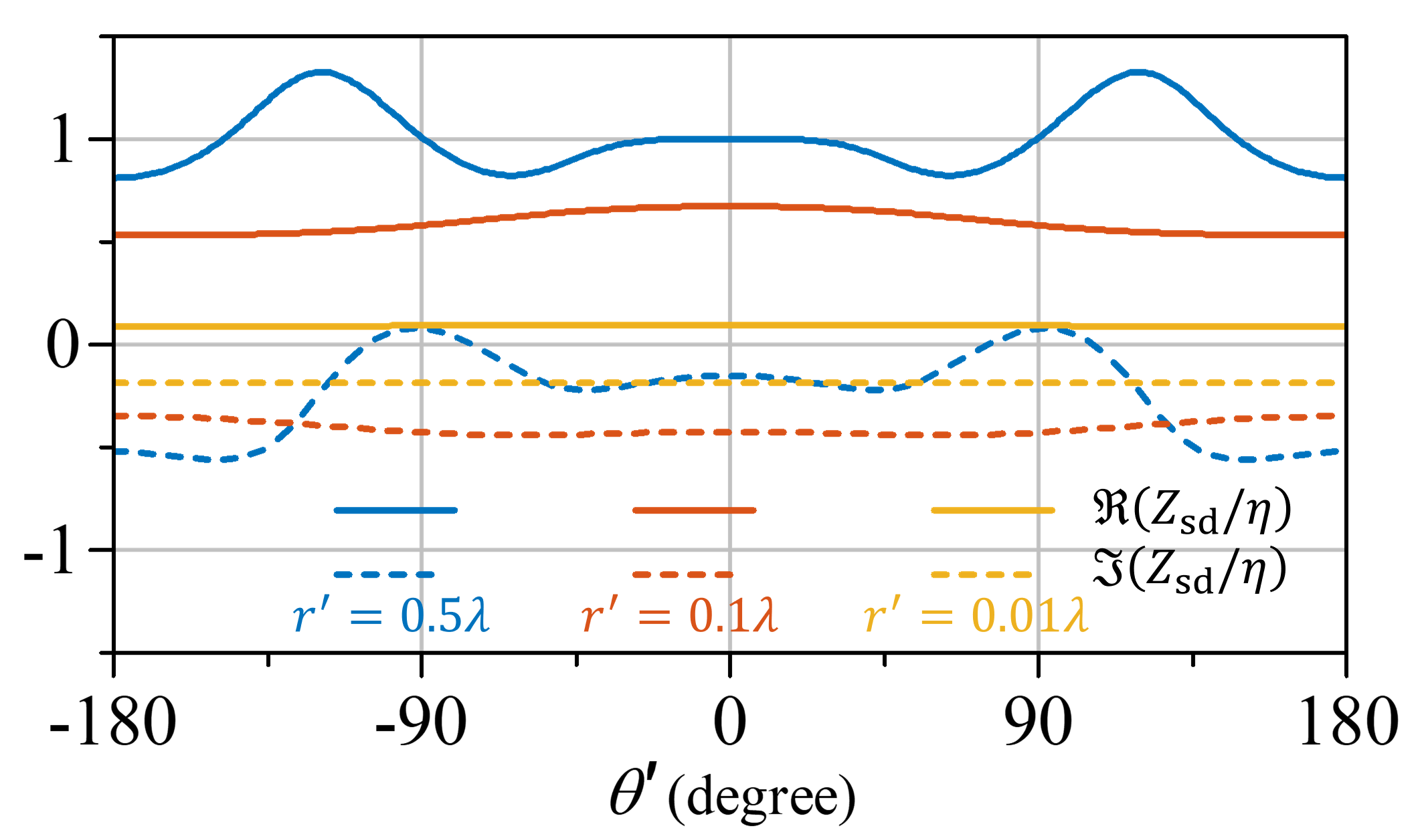}\label{fig:n3b}}
    \caption{Numerical simulation of a cavity metamirror that focused the fields generated by a point source into an image point inside the cavity volume. (a) The total electric field and Poynting vector (represented by the black arrows). (b) The required surface impedance to implement a perfect drain with different radii.}
    \label{fig:n3}
\end{figure}

However, preceding the study of the applicability of power flow-conformal metamirrors for focusing, it is important to discuss the role of the virtual source generating the reflected fields. 
Mathematically, the virtual source is a time-reversed version of the actual source whose role is to extract the energy introduced in the system by the actual source. In other words, the virtual source acts as a drain that totally absorbs all incident radiation without reflections. 
For its implementation, one may consider an active, phase-retarded perfect drain that coincides with an inward monopole wave and is designed to cancel input signals \cite{de2002overcoming,leonhardt2009perfect}. Even though active drains can be in principle realized, their implementation requires well-controlled temporal and spatial synchronization \cite{kinsler2010active}.
The conditions for passive perfect drains have been discussed in the literature where we can find solutions based on engineering the complex permittivity of non-magnetic materials \cite{gonzalez2011perfect} or the use of loaded coaxial cables \cite{sun2010can}.
It is important to mention that, as it happens in the case of Maxwell’s fish-eye lens bounded by a perfectly conducting (PEC) boundary, the focusing functionality with super-resolution is only achieved in the presence of the drain, so it is important to extend the analysis and understanding of this singular point.    

We start the analysis of the drain by considering a cylindrical surface surrounding the singular line (point in 2D space) [see Fig.~\ref{fig:n3a}]. The boundary conditions at this cylindrical boundary can be defined by the surface impedance. Fig.~\ref{fig:n3b} represents both the real and imaginary parts of the surface impedance for three different surfaces with different radii: $r'=0.5\lambda,\, 0.1\lambda,\, 0.01\lambda$. We can see that in these three cases the real part of the surface impedance is positive, as the drain behaves as a perfectly matched absorber. The variation of the surface impedance around the cylindrical surface of the drain is reduced when the radius of the drain is reduced, which is due to the weaker variation of the incident field in a smaller circle: very close to the focus point the total field is dominated by the field of the converging cylindrical wave, which depends only on the distance from the focal point.

We can see that both real and imaginary parts of the surface impedance tend to zero when the radius of the drain tends to zero.
However, in the limit of zero radius, the physical meaning of the surface impedance loses its sense, as the two-dimensional cylindrical surface shrinks to a one-dimensional line. In this limit,  we should use the model of an impedance wire whose parameter is the impedance per unit length. 
This impedance is defined as the ratio of the longitudinal component of the electric field at the wire surface to the total current flowing along the wire which is equal to the circulation of the magnetic field around the wire. Because in the very vicinity of the sink the field is very close to that of the sink in free space, we can find this impedance directly from Eq.~\eqref{E_sink}:
\begin{equation}
    Z_{\rm pul}=\eta{k\over 4}+j\eta{k\over 2\pi}(1+\ln {{\gamma kr_{\rm d}\over 2} }).\label{Zpul}
\end{equation}
Here we have used the small-argument approximation of the Hankel function, 
$H_0^{(1)}(x)\approx 1+j{2\over \pi}(1+\ln {\gamma x \over 2})  $, where $\gamma\approx 0.5772$ is the Euler constant. Alternatively, we can calculate the wire impedance per unit length as 
\begin{equation}
  Z_{\rm pul}=\lim_{r_{\rm d}\rightarrow 0}{E_z\over 2\pi r_{\rm d} H_\phi}  
\end{equation}
which gives the same result. Indeed, the amplitude of the $\phi$-component of the magnetic field reads, from \eqref{E_sink},  $H_\phi=jI_0{k\over 4}H_1^{(1)}(kr_{\rm d})$. Using the small-argument approximation $H_1^{(1)}(x)\approx {x\over 2}-j{2\over \pi x} $ we arrive to the same  result \eqref{Zpul}. Note that the real part of this impedance equals the radiation resistance (per unit length) of a radiating current line, as it should be. The wire reactance is capacitive (because $kr_{\rm d}\ll 1$), and it is the negative of the inductive reactance per unit length of a thin conducting wire acting as a power source, indicating a kind of resonant condition of the whole structure.

Figure~\ref{fig:n3a} shows the results of  numerical simulations of a cylindrical metamirror cavity whose radius is $6\lambda$ [surface impedance for this boundary,$Z_{sc1}$, is represented in Fig.~\ref{fig:n2c}]. The radius of the drain is $r'=0.1\lambda$. Here we can see how the metamirror cavity is able to produce a converging cylindrical wave at the position of the drain. Black arrows represent the total Poynting vector and show how the energy is emerging from the source and converging into the drain.

\begin{figure}[t]
    \centering
    %\vspace{-1cm}
    \subfigure[]{\includegraphics[width=0.65\linewidth]{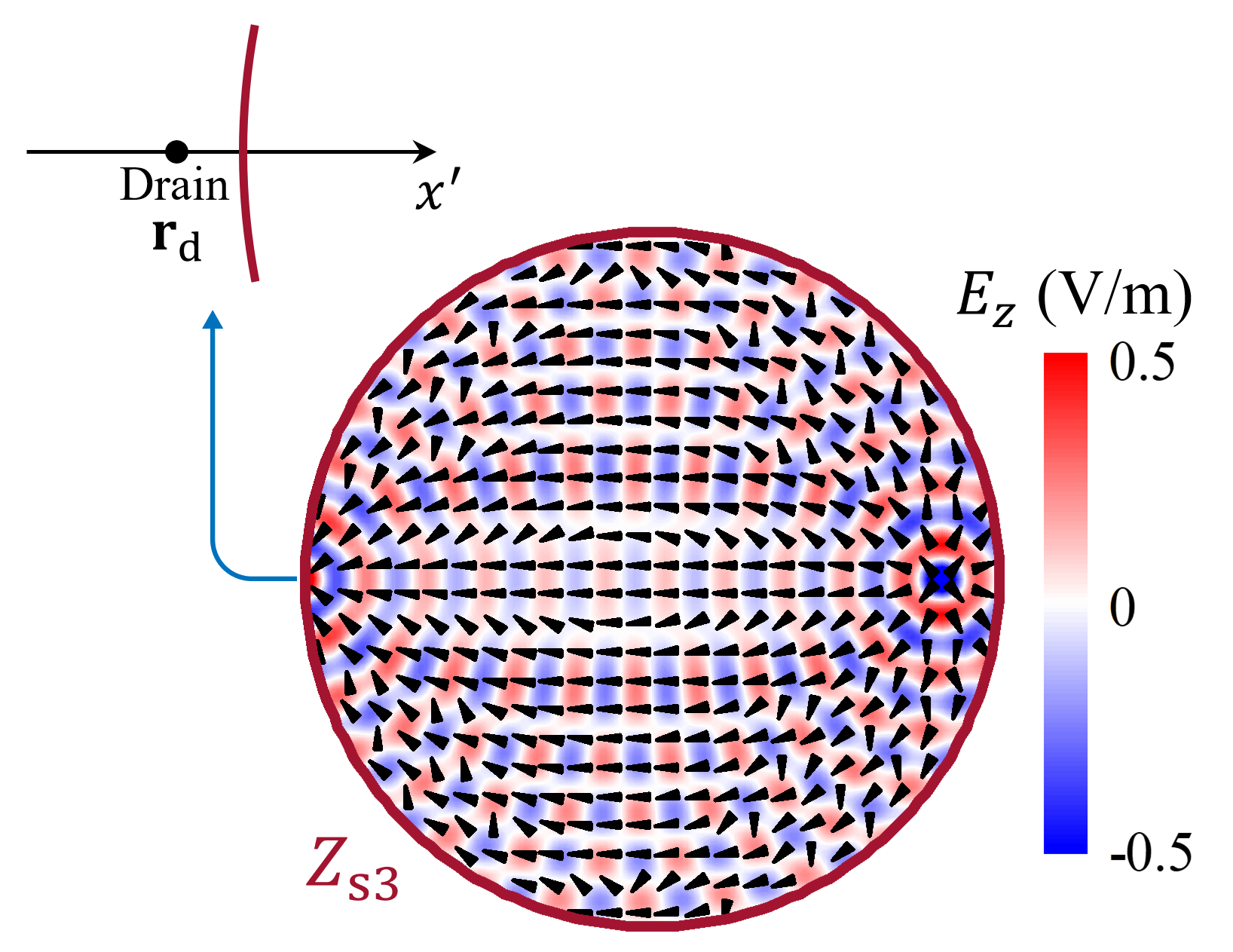}\label{fig:n4a}}
    \subfigure[]{\includegraphics[width=0.8\linewidth]{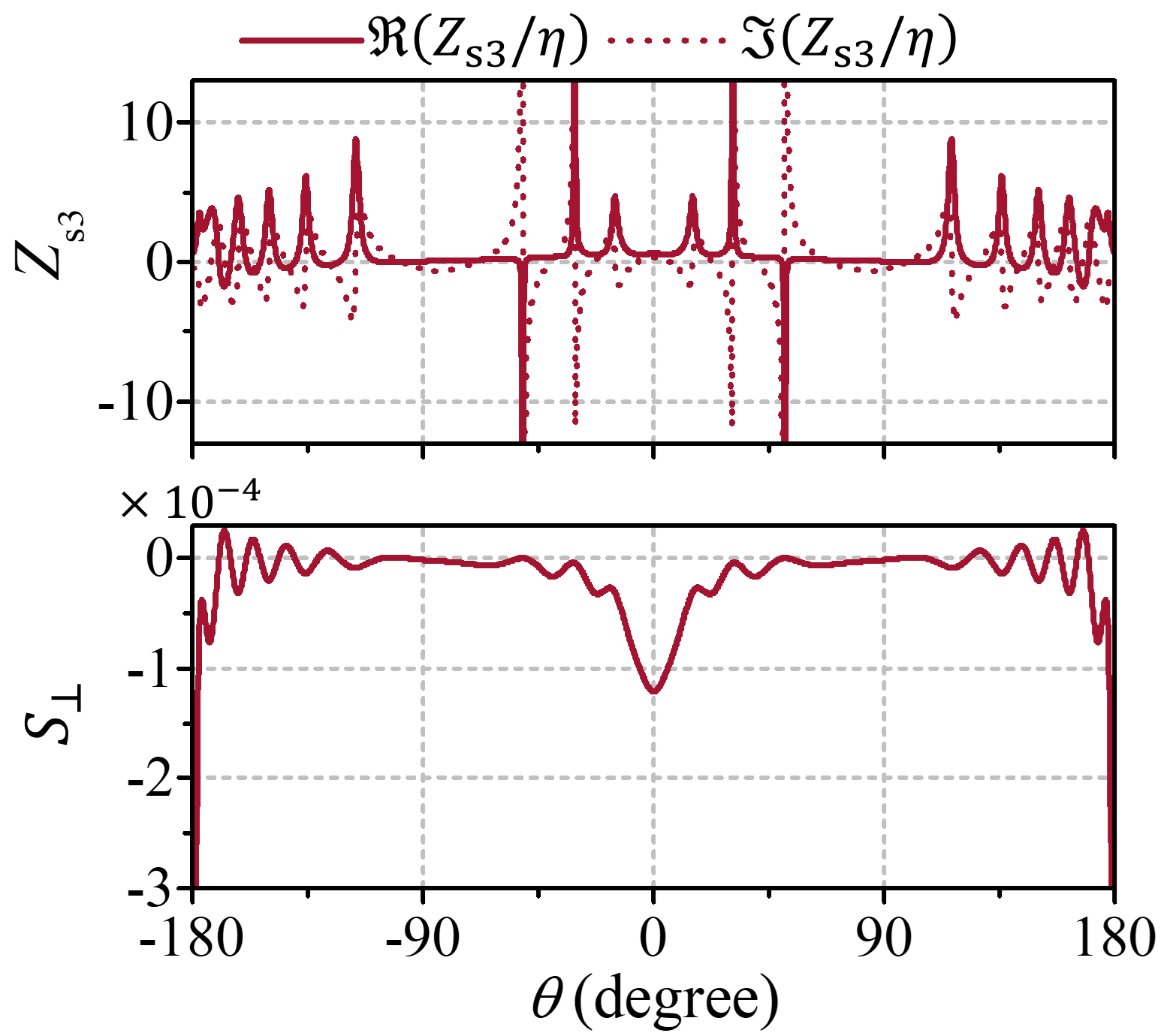}\label{fig:n4b}}
    \caption{Numerical simulation of a cavity metamirror that focuses the fields generated by a point source into an image point outside of the cavity volume. (a) The total electric field and Poynting vector (represented by the black arrows). (b) The required surface impedance to implement the cavity and the normal component of the Poynting vector at the cavity surface.}
    \label{fig:n4}
\end{figure}

An alternative possibility is to place the drain outside of the closed surface, as shown in  Fig.~\ref{fig:n4}. As an example, Fig.~\ref{fig:n4a} shows the results of numerical simulations of a cylindrical cavity with the radius $6\lambda$ when the source is located at $\rvs=(5\lambda, 0)$ and the drain is defined at $\rvd=(-6.1\lambda, 0)$ (outside of the cylindrical cavity). The surface impedance required to implement the cavity with this configuration is shown in Fig.~\ref{fig:n4b}. As in the previous example, the surface impedance is ``active-lossy". However, in this scenario, the overall response of the metamirror is not lossless. The metamirror itself is acting as a drain and absorbs the energy introduced in the system by the source. This behavior can be seen in the plot of the local power entering and emerging from the metamirror [bottom panel in Fig.~\ref{fig:n4b}], where we can see how the loss and gain are distributed along the metamirror. This result demonstrates that the drain (power receiver) can be implemented as an independent element inside the cavity or integrated into the surface of the cavity.

\begin{figure*}

\minipage{0.23\textwidth}
    \subfigure[]{\includegraphics[width=1\textwidth]{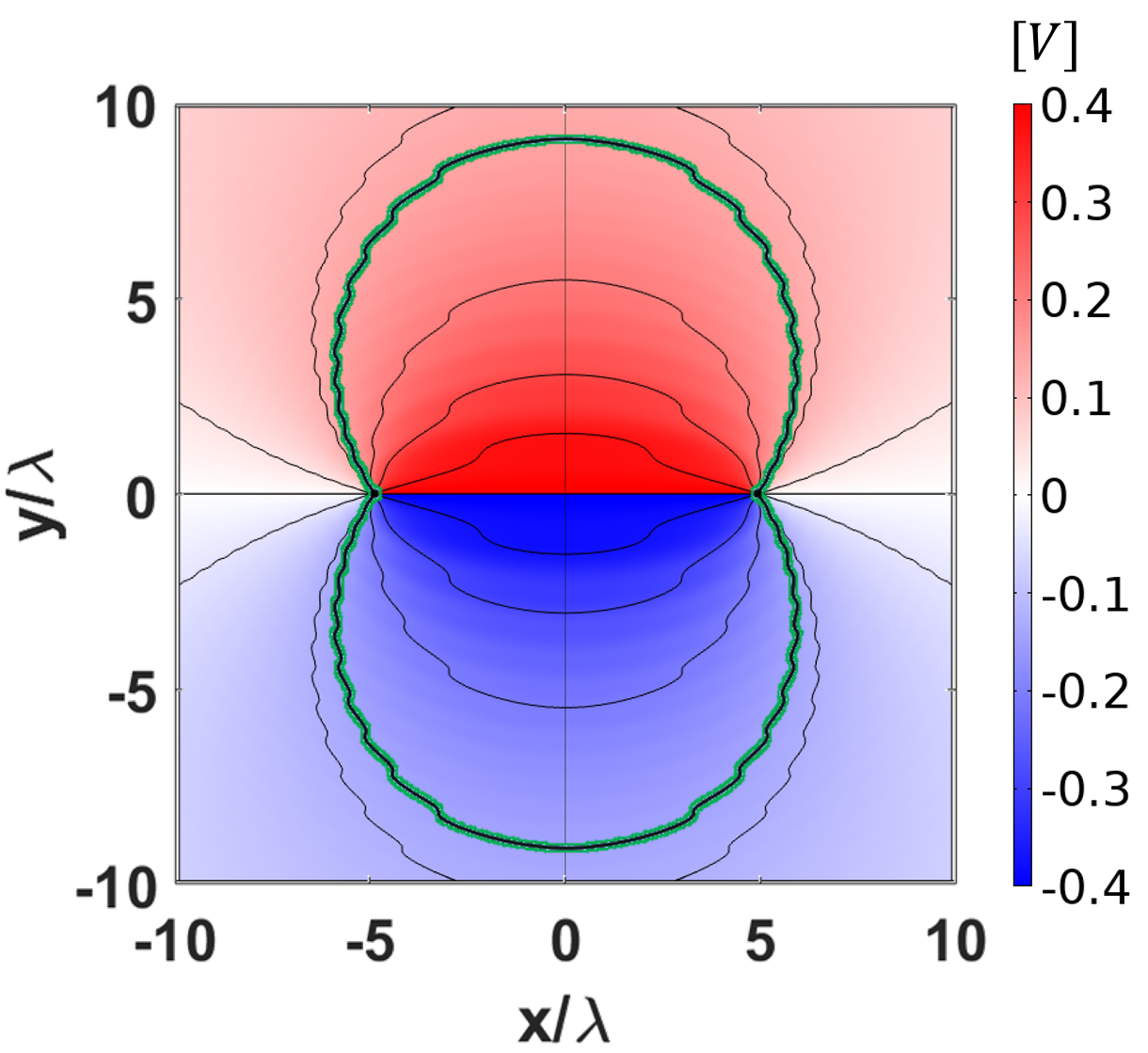}  \label{fig:n5a}}
\endminipage\hfill
\minipage{0.27\textwidth}
    \subfigure[]{\includegraphics[width=.9\textwidth]{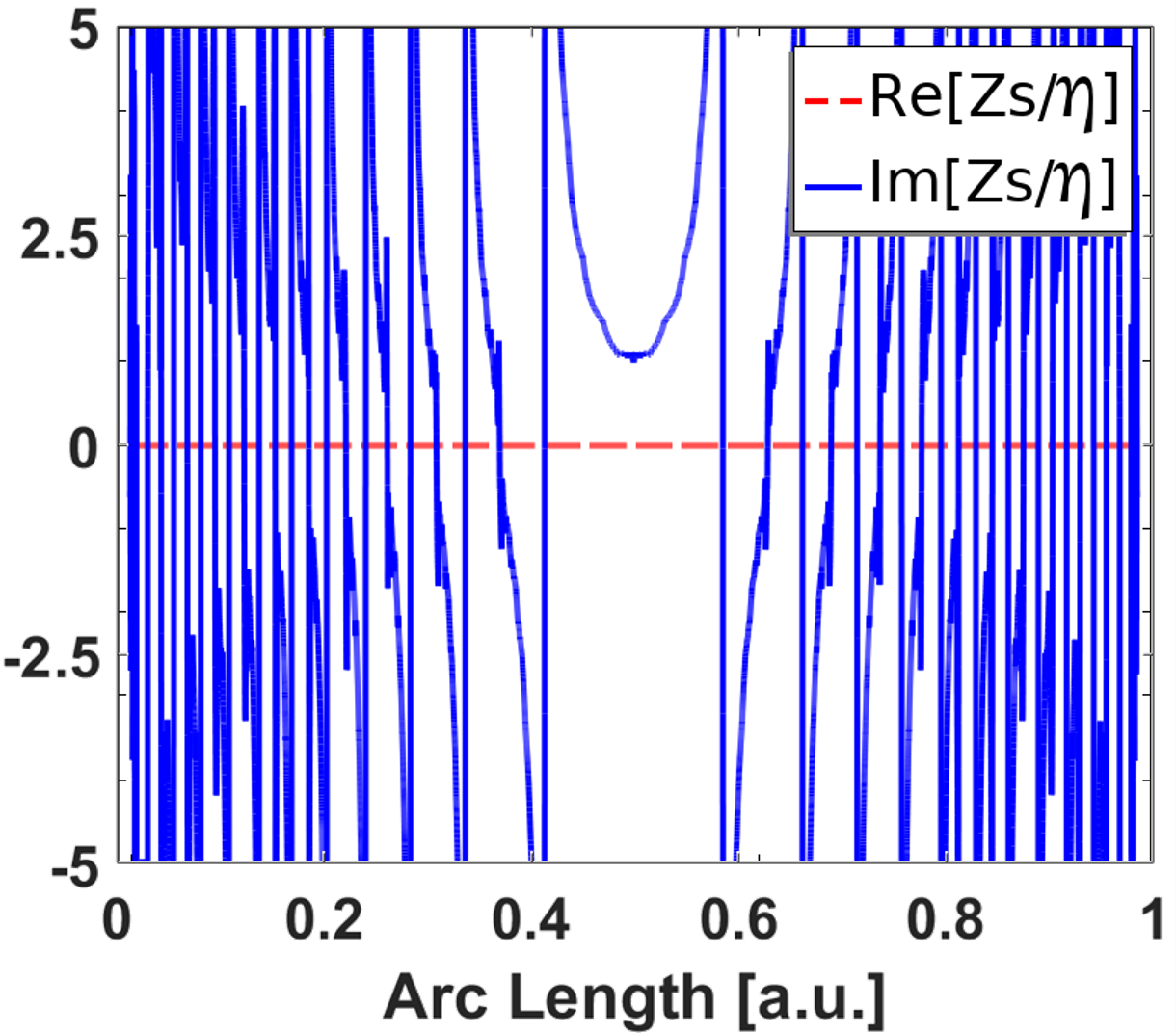}\label{fig:n5b}}
\endminipage\hfill
\minipage{0.21\textwidth}
    \subfigure[]{\includegraphics[width=1\textwidth]{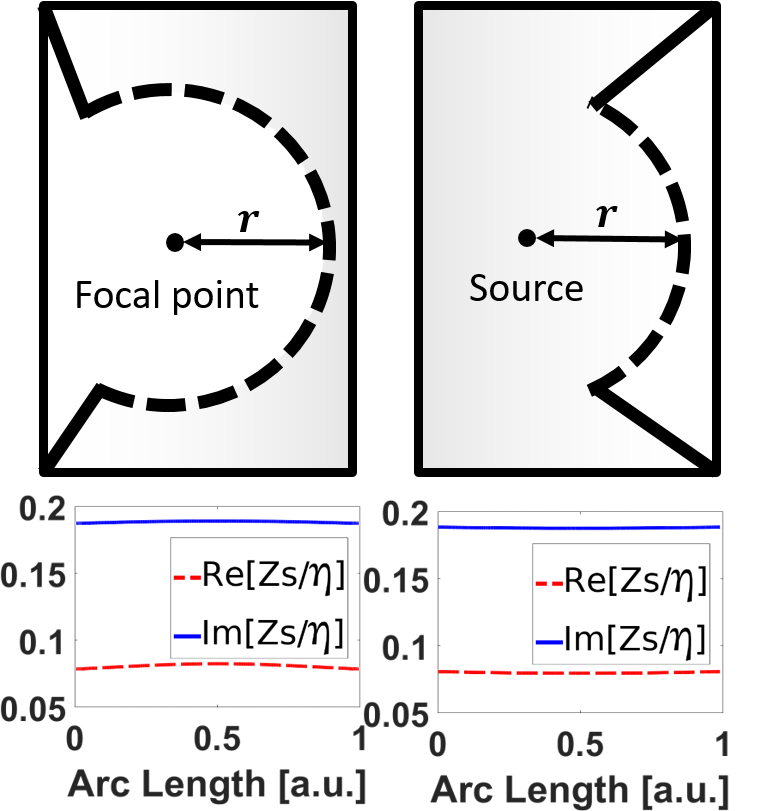} \label{fig:n5c}}
\endminipage\hfill
\minipage{0.26\textwidth}
    \subfigure[]{\includegraphics[width=1\textwidth]{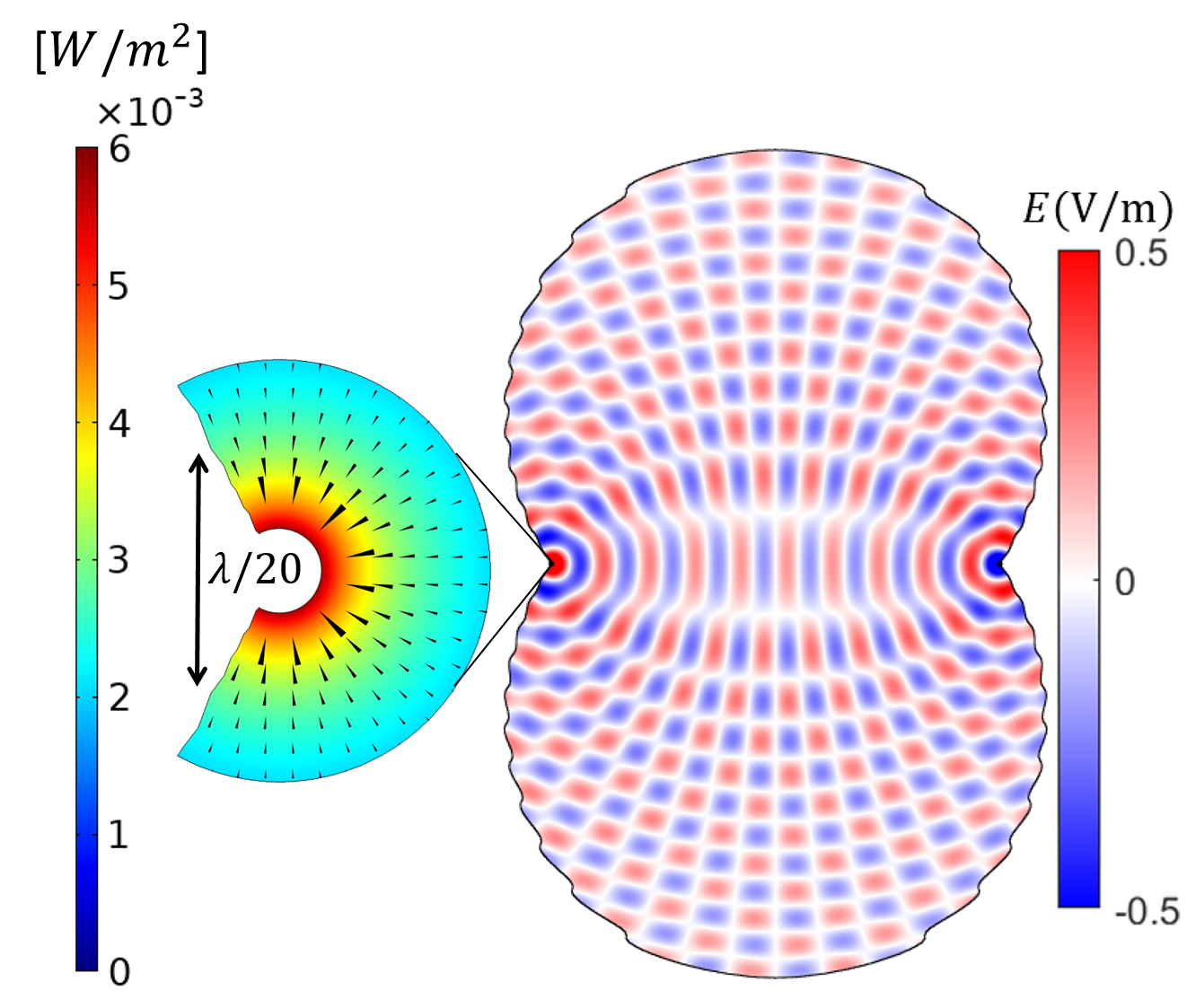}\label{fig:n5d}}
\endminipage\hfill

    \caption{(a) Electrostatic study to find the shape of power-conformal surfaces. Black lines represent a set of conformal surfaces which are tangential to the Poynting vector. (b) Normalized surface impedance for the highlighted conformal surface (green color) in (a). (c) Open boundaries are closed with lossy arcs (the part of a circle with radius $r=0.01\lambda$). Solid lines are power-conformal, and dashed lines are lossy boundaries. (d) Simulation results of the subwavelength focusing by the conformal cavity closed with circular arcs.}
   \label{fig:3}
 \end{figure*}

Now we are ready to consider the possibility to deform the surface into a power-conformal shape so that the surface will be lossless at all points. 
To this end, we define a vector perpendicular to the Poynting vector as $\_N=-S_y\hat{x}+S_x\hat{y}$ and, using the properties of the gradient, we look for a function $g(x,y)$ such that $\nabla g(x,y)=\_N$. 
This expression resembles the well-known formula for the electric potential in electrostatics, $E=-\nabla V$.
%and we can calculate $g(x, y)$ as an electrostatic problem. 
Using this mathematical analogy, we calculate $g(x, y)$ using an electrostatic solver in COMSOL. By defining the electric displacement field to be proportional to vector $\_N$ we solve the electric potential that is equivalent to the function $g(x,y)$ [see Fig.~\ref{fig:n5a}]. 
Finally, the profile of the metamirror can be defined as an equipotential surface.
There are infinitely many surfaces tangential to the power flow, and each of them defines a different metamirror with different electromagnetic responses.
Black lines in Fig.~\ref{fig:n5a} graphically represent a set of conformal surfaces which are tangential to the Poynting vector of the desired field distribution. It is important to notice that, in order to calculate the electric potential, the singularities associated with the source and the drain have to be excluded from the simulation domain, that is, we do not obtain a closed surface. The definition of the electromagnetic properties around these singular points will be discussed later.

To engineer a metamirror based on this analysis, we select one of the conformal lines and calculate the surface impedance defined by Eq.~\eqref{Eq:Zs}. Figure~\ref{fig:n5b} shows the surface impedance for the highlighted conformal surface (green color).   
%The power flow-conformal surfaces sketched with solid blue and red lines. 
The surface impedance is represented for one section, as it is symmetric with respect to the $y$-axis. The real part of the surface impedance is zero, indicating the local and lossless response of the metamirror.
%the fact that selected surface is power-conformal indeed. 
The imaginary part of the impedance shows the rapid fluctuation of the reactance, responsible for wavefront engineering. 
%Similar behavior is reported with other methods \cite{9119196}.

As we have mentioned, the system has two field singularities at the source and focal points. Therefore, close vicinity areas were excluded from the electrostatic analysis, so the equipotential lines do not build closed surfaces.  This leaves the boundary open around both singularities. Although open boundaries are very small, in the proximity to the source and the drain nearly all the power crosses these boundaries.  
In our approach, we complete the surface with circular arcs [see Fig.~\ref{fig:n5c}]. 
These active or lossy surfaces are shown by dashed arcs with the radius $r=0.01\lambda$ in Fig.~\ref{fig:n5c}. Normalized surface impedances of the corresponding circular arc boundaries are drawn in the same picture. In both cases, the real parts of the surface impedance are positive indicating the presence of loss. So, one can realize this surface impedance with passive elements. In the current scenario, the source is inside the simulation domain and the focal point is excluded from the cavity domain, so the metasurface is overall lossy and all the losses are concentrated in the closing arcs. This scenario corresponds to placing the receiver just at the mirror surface.

Figure~\ref{fig:n5d} shows the results of a numerical simulation of the conformal cavity closed by circular arcs. This simulation shows the total field generated by the point source, which is in excellent agreement with the theoretical results.  We also show  the absolute value of the reflected power $P_r=\frac{1}{2\eta}|E_r|^2$ $[W/m^2]$ from the metamirror. The power distribution shows that almost all of the input power is focused at the focal point. In this particular design, we observe subwavelength focusing with the hotspot size $HBPW=\lambda/20$, and it is limited only by the excluding circular arc $r=0.01\lambda$. In theory, this is the proof of concept for perfect focusing, because by implementing a smaller arc it is possible to further reduce the hotspot size.  
%Also, rough edges at the interface of open boundary and reflector, create unwanted scattering fields. So, this is a challenge to be solved toward perfect focusing.

%\begin{equation}
%  \begin{split}
%E_{{\rm tot}z}=\textcolor{red}{E_0\left[J_0(kr_{\rm d})+J_0(kr_{\rm s})+j\left[Y_0(kr_{\rm d})-Y_0(kr_{\rm s})\right]\right]}
%\label{eq:Ez}
%  \end{split}
%\end{equation} 
%\begin{equation}
%  \begin{split}
%H_{{\rm tot}x}=\textcolor{red}{\frac{E_0}{\eta}\left[\frac{y-y_{\rm d}}{r_{\rm d}}Y_1(kr_{\rm d})-\frac{y-y_{\rm s}}{r_{\rm s}}Y_1(kr_{\rm s})\right]}\\
%\textcolor{red}{-j\frac{E_0}{\eta}\left[\frac{y-y_{\rm d}}{r_{\rm d}}J_1(kr_{\rm d})+\frac{y-y_{\rm s}}{r_{\rm s}}J_1(kr_{\rm s})\right]}
%\label{eq:Ez}
%  \end{split}
%\end{equation} 

%where $E_0=\frac{I_0\eta k}{4}$, $J_0$ and $J_1$ are the Bessel functions of the first kind of orders $0$ and $1$, respectively. Also, $Y_0$ and $Y_1$ are the Bessel functions of second kind of orders $0$ and $1$, respectively. Figure \ref{fig:n2a}

{\bf

}

\section{Open metamirrors for subwavelength focusing}
\label{sec:open}

\begin{figure*} 
\minipage{0.2\textwidth}
    \subfigure[]{\includegraphics[width=1\textwidth]{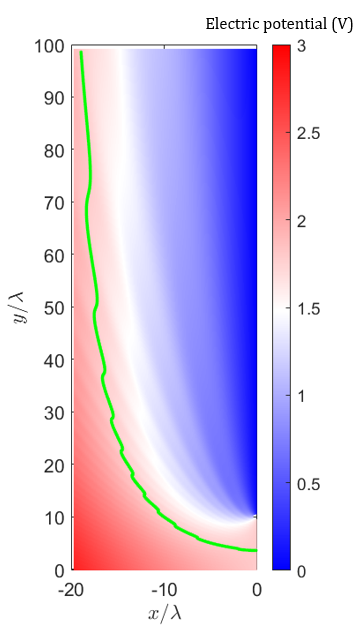}  \label{fig:n6a}}
\endminipage\hfill
\minipage{0.3\textwidth}
    \subfigure[]{\includegraphics[width=.9\textwidth]{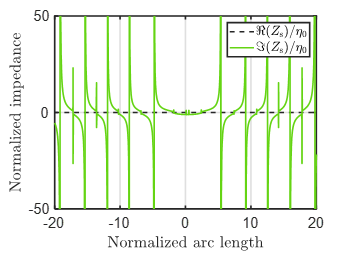}  \label{fig:n6b}}
    \subfigure[]{\includegraphics[width=.9\textwidth]{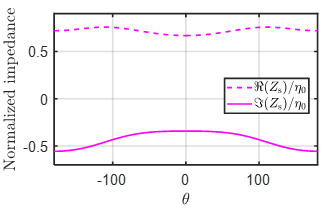}  \label{fig:n6c}}
\endminipage\hfill
\minipage{0.45\textwidth}
    \subfigure[]{\includegraphics[width=1\textwidth]{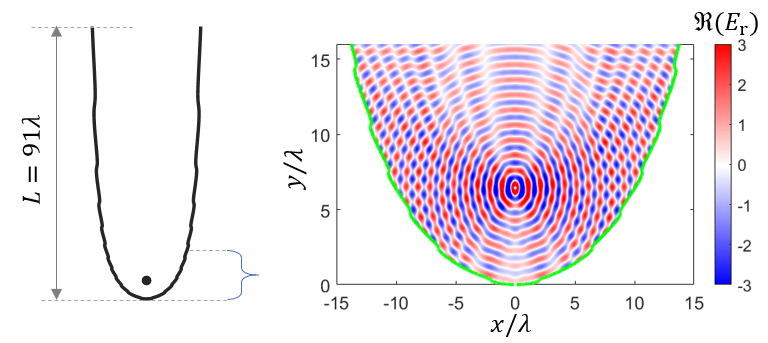}  \label{fig:n6d}}
    \subfigure[]{\includegraphics[width=1\textwidth]{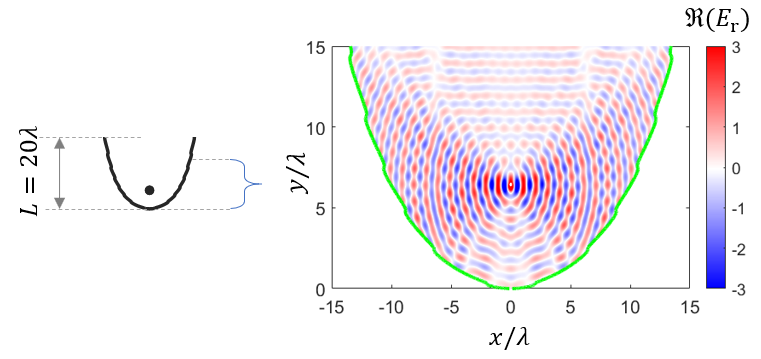}  \label{fig:n6e}}
\endminipage\hfill
    \caption{(a) Electrostatic study to find power-conformal surfaces when the amplitude of the incident plane wave is $E_0=1$~V/m, the aperture of the reflector is $D=40\lambda$, and the position of the virtual source is $\rvd=(0, 10\lambda)$. The green line represents a conformal surface that is tangential to the Poynting vector. (b) Normalized surface impedance for the highlighted conformal surface (green color) in (a). (c) Normalized surface impedance for implementing a prefect drain by a circular surface with the radius $r'=0.1\lambda$. (d) Numerical simulation of the reflected field scattered by the conformal metamirror with a total length $L=91\lambda$. (e) Numerical simulation of the field reflected by the conformal metamirror with the total length $L=20\lambda$. The real part of the electric fields around the focal point is represented in (d) and (e).
}
   \label{fig:6}
 \end{figure*}
 
In this section, we use the same methodology for studying open metamirrors to focus the field of an impinging plane wave. In this case, the incident fields can be expressed as
%\begin{eqnarray}
%E_{{\rm s}z}=E_0 e^{-jk(\sin\theta_{\rm i}x-\cos\theta_{\rm i}y)}\\
%H_{{\rm s}x}=-\frac{\cos\theta_{\rm i}}{\eta}E_0 e^{-jk(\sin\theta_{\rm i}x-\cos\theta_{\rm i}y)} \\
%H_{{\rm s}y}=-\frac{\sin\theta_{\rm i}}{\eta}E_0 %e^{-jk(\sin\theta_{\rm i}x-\cos\theta_{\rm i}y)} \
%\label{eq:inc_pw}
%\end{eqnarray}  
\begin{equation}
 \Es=E_0 e^{-jk(\sin\theta_{\rm i}x-\cos\theta_{\rm i}y)}\hat{z}, \hspace{5mm}
	\Hs = \frac{j}{\eta k}\D\x\Es,
	\label{eq:inc_pw}
\end{equation}
where $\theta_{\rm i}$ is the angle of incidence. The field of the virtual source generating the reflected fields is defined, as in the previous example by Eq.~(\ref{E_sink}). In this case, the amplitude of the virtual source has to be adjusted to make sure that the total reflected power equals the incident power. If we assume that the reflection plane is normal to the $y$-direction,  and the incident wave propagates along the normal direction ($\theta_{\rm i}=0$), the power introduced in the system can be calculated as $P_{\rm pw}=\frac{1}{2\eta}DE_0^2 $ with $D$ being the aperture of the metamirror. Then, the virtual source defines the reflected power as $P_{\rm d}=\frac{1}{2}\Re(Z_{pul})I_0^2 $. Equating these two values we obtain the amplitude of the current in the virtual source as a function of the amplitude of the plane wave and the effective aperture of the metamirror as
\begin{equation}
    I_0=\frac{2}{\eta}E_0\sqrt{\frac{D}{k}}.
\end{equation}

Following a similar approach to that in Section~\ref{sec:cavity}, the first step is to perform the electrostatic study and find the shape of power-conformal surfaces.  In particular, we will consider an incident plane wave with the amplitude $E_0=1$~V/m, a reflector with the aperture $D=40\lambda$, and the position of the virtual source at $\rvd=(0, 10\lambda)$. Defining the electric displacement to be proportional to the vector $\_N$ (a vector perpendicular to the Poynting vector), we obtain the electric potential represented in Fig.~\ref{fig:n6a}. In this calculation, we exclude a small domain near the drain and exploit the symmetry of the problem with respect to the $y$-axis.  As it has been discussed, the profile of the power flow-conformal metamirror can be defined as an equipotential surface. In this case, among all the possible surfaces, we are interested in a surface that closes behind the drain (shown in green color). Notice, that this solution is unique and other surfaces will converge into the source or diverge to infinity. 

It is interesting to see how the surface of the conformal metamirror becomes more vertical and tends to the aperture of the metamirror $D$ defined in the design process. This shape of the conformal surface ensures that the energy entering the system is in agreement with the power balance stipulated by the design conditions.    
Next, in order to implement the metamirror, we need to calculate the surface impedance on this conformal surface.  Figure~\ref{fig:n6b} represents the normalized surface impedance where we can see that, as expected, the surface impedance is purely imaginary at all points.  

To verify the properties of the designed surface, we numerically calculate the scattered fields. In this calculation, we implement the drain as a lossy surface impedance on a circular surface around the drain point, with the radius $r'=0.1\lambda$. The surface impedance of the drain is represented in Fig.~\ref{fig:n6c}. Before we start with the analysis, it is important to mention that the power-conformal surface obtained from the analysis of the power flow is infinite and, in a practical scenario, we need to truncate it and define a finite length, $L$. Obviously, the scattering properties of the metamirror will depend on the mirror length, and we will study this effect by studying two examples.  
The first example is a metamirror with the total length of $L=91\lambda$. In this case, the aperture of the metamirror is $D\approx37\lambda$  (close to the design conditions). The reflected field near the drain is shown in Fig.~\ref{fig:n6d} where we can see that the reflected field focuses on the drain position. The pattern of the cylindrical wave is perturbed due to parasitic reflections caused by the finite size of the reflector. 
For comparison, we simulate a metamirror with the same configuration but with a length $L=20\lambda$. The reflected fields are shown Fig.~\ref{fig:n6e}. In this example, we can see how the waves still focus on the drain, but there is a deterioration of the cylindrical pattern as a consequence of a smaller reflector size.

\begin{figure}[]
 \subfigure[]{\includegraphics[width=0.48\linewidth]{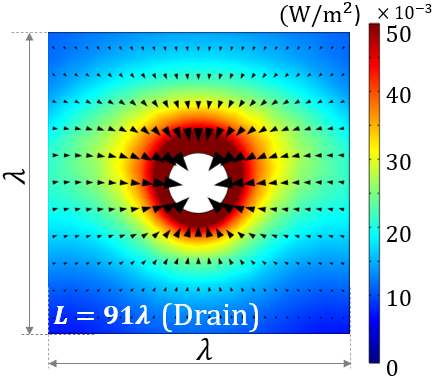}  \label{fig:7a}}
 \subfigure[]{\includegraphics[width=0.48\linewidth]{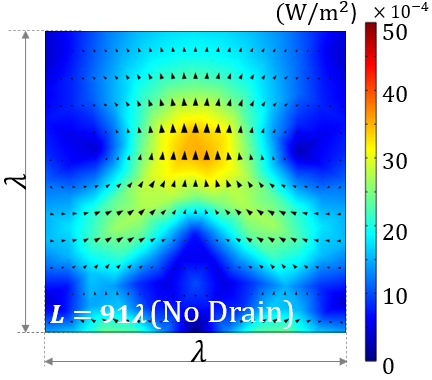}  \label{fig:7b}}
  \subfigure[]{\includegraphics[width=0.48\linewidth]{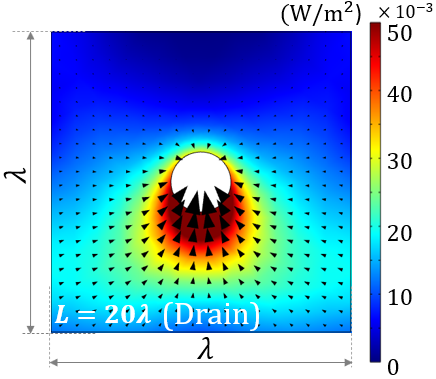}  \label{fig:7c}}
 \subfigure[]{\includegraphics[width=0.48\linewidth]{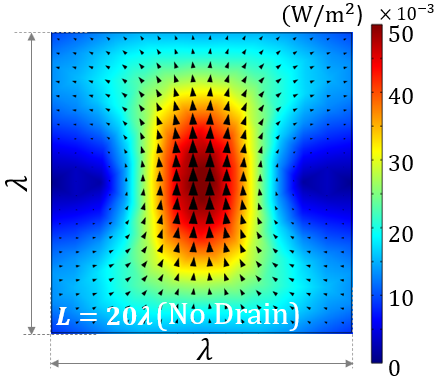}  \label{fig:7d}}
 \caption{Effect of the drain in the power localization. (a) The absolute value of the time-averaged power when the conformal metamirror has the total length $L=91\lambda$ and the prefect drain is implemented by a circular surface with the radius $r'=0.1\lambda$. (b) The absolute value of the time-averaged power when the conformal metamirror has the total length $L=91\lambda$ without a drain. (c) The absolute value of the time-averaged power when the conformal metamirror has the total length $L=20\lambda$ and the prefect drain is implemented by a circular surface with the radius $r'=0.1\lambda$. (d) The absolute value of the time-averaged power when the conformal metamirror has the total length $L=20\lambda$ without a drain. Black lines represent the power flow.}
\label{fig:7}
\end{figure}

Further analysis is required to understand the role of the drain on power localization. Using the same examples shown in Fig.~\ref{fig:6}, we study the power localization in the presence and in absence of the drain. Firstly, we study the metamirror with the total length $L= 91\lambda$ and a perfect drain implemented by a circular surface with the radius $r'=0.1\lambda$.  Figure~\ref{fig:7a} shows the time-averaged power flow near the drain where we can see how the drain acts as a sink and collects the power. In this case, the size of the focusing spot depends on the physical size of the drain. Figure~\ref{fig:7b} shows the power flow for the same metamirror without a drain. We can see that there is strong localization at the focal point. However,  in the absence of drain, the energy of the converging cylindrical wave is scattered back (to satisfy the energy conservation), and the field strength at the focal point is reduced. In the ideal case of an infinite large metamirror that produced a perfectly converging cylindrical wave, without a drain, there would be no power localization.  Figure~\ref{fig:7c} shows the power localization produced by a metamirror with the total length $L= 20\lambda$ and a perfect drain implemented by a circular surface with the radius $r'=0.1\lambda$.  Here, we can see how the energy converges to the drain but the distortion of the cylindrical pattern as a consequence of the parasitic reflections creates an asymmetric distribution of the power around the drain (most of the power is absorbed from the bottom side of the drain). The results for the same structure without the drain are shown in Figure~\ref{fig:7d}. In this case, the power localization is stronger than in Fig.~\ref{fig:7b}, but the size of the focal spot is limited by diffraction.

\begin{figure}[]
 \subfigure[]{\includegraphics[width=0.8\linewidth]{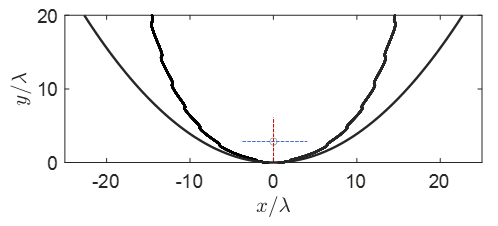}  \label{fig:8a}}
 \subfigure[]{\includegraphics[width=0.8\linewidth]{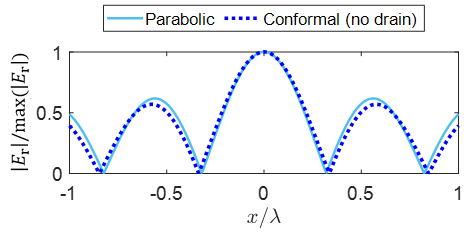}  \label{fig:8b}}
  \subfigure[]{\includegraphics[width=0.8\linewidth]{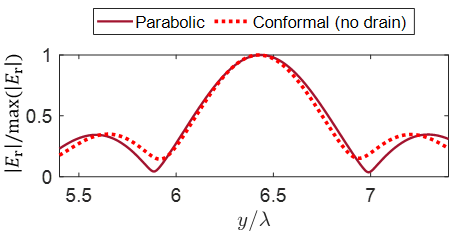}  \label{fig:8c}}  
 \caption{Comparison between a parabolic reflector and a power-conformal metamirror without a passive drain. (a) The surface of a parabolic reflector and the conformal metamirror. (c, d) Comparison of the normalized electric field scattered by the two structures in the $x$-plane (b) and $y$-plane (c) at the focal point.}
\label{fig:8}
\end{figure}

To finalize this study, we compare the conformal metamirror without a drain and a conventional parabolic reflector. The surface profiles for the two structures are presented in Fig.~\ref{fig:8a}. To compare the results for both structures, we study the electric field distribution in two orthogonal planes in the vicinity of the focal point. Figures~\ref{fig:8b} and \ref{fig:8c} represent the field distribution in the $x$-plane and $y$-plane, respectively. Notice that the amplitude of the electric field is normalized to the maximum value at the focal point. We can see that in the absence of a passive drain (power receiver) the power-conformal mirror offers similar focusing performance as a conventional parabolic mirror. However, it is important to stress that the aperture of the power-conformal metamirror is significantly smaller than that of the corresponding parabolic mirror.

\section{Conclusion}
\label{sec:con}
Based on the power flow analysis, we have established a method to engineer metamirrors for point-to-point subwavelength focusing. Compared to genuine geometry designs such as a flat reflector or a circular cavity, introduced metamirrors are shaped along with the power flow of the desired power distribution. They can be implemented by lossless elements. To deplete the input power of the impinging waves out of the system, lossy elements at or around the focal point are required. We have analyzed the impact of the drain whether it is inside or outside the cavity and found the corresponding surface impedance. We have shown that when the drain is outside the cavity, the surface is no longer lossless, but the power absorption can be concentrated at an arbitrarily small area of the surface. In this case, the radius of the drain arc defines the hotspot size of the focusing device. For instance, with the radius $r=0.01\lambda$, we have achieved subwavelength focusing with $HPBW=\lambda/20$. This is a theoretical proof of the concept for perfect focusing, as it is possible to obtain even smaller hotspot sizes within the limit of smaller drains. 

Using the same approach, we designed open metamirrors to focus incident plane waves. With the presence of a drain at the focal point, we have shown subwavelength focusing also for open power-conformal reflectors.  Compared to a parabolic reflector in the absence of drain,  engineered power-conformal metamirrors can provide the same quality of focusing with much smaller apertures. We hope that the results of this work pave the way for developing applications that require subwavelength focusing.

\begin{acknowledgments}
This work has been supported by the European Commission under grant H2020-FETOPEN-736876 (VISORSURF). 
\end{acknowledgments}

%\appendix

%\section{Appendixes}

%\nocite{*}
%\bibliographystyle{h-physrev}
\bibliography{Library}% Produces the bibliography via BibTeX.

\end{document}